\documentclass[aip,reprint]{revtex4-1}

\usepackage{graphicx}
\usepackage{epsfig}
\usepackage[pdfversion=1.3,pdfversion=1.4]{epspdfconversion}

\draft 

\begin{document}


\title{Sub 20\,nm Silicon Patterning and Metal Lift-Off Using Thermal Scanning Probe Lithography}

\author{H. Wolf}
\author{C. Rawlings}
\author{P. Mensch}
\affiliation{
IBM Research-Zurich, S\"aumerstrasse 4,
8803 R\"uschlikon, Switzerland
}%

\author{ J. L. Hedrick}
\author{ D. J. Coady}
\affiliation{%
IBM Research-Almaden, 650 Harry Road, San Jose, CA 95120, USA
}
\author{U. Duerig}
\affiliation{
IBM Research-Zurich, S\"aumerstrasse 4,
8803 R\"uschlikon, Switzerland
}%
\author{A.W. Knoll}%
 \email{ark@zurich.ibm.com.}
\affiliation{
IBM Research-Zurich, S\"aumerstrasse 4,
8803 R\"uschlikon, Switzerland
}%

\date{\today}

\begin{abstract}
The most direct definition of a patterning process' resolution is the smallest half-pitch feature it is capable of transferring onto the substrate. Here we demonstrate that thermal Scanning Probe Lithography (t-SPL) is capable of fabricating dense line patterns in silicon and metal lift-off features at sub 20\,nm feature size. The dense silicon lines were written at a half pitch of 18.3\,nm to a depth of 5\,nm into a 9\,nm polyphthalaldehyde thermal imaging layer by t-SPL. For processing we used a three-layer stack comprising an evaporated SiO$_2$ hardmask which is just 2-3\,nm thick. The hardmask is used to amplify the pattern into a 50\,nm thick polymeric transfer layer. The transfer layer subsequently serves as an etch mask for transfer into silicon to a depth of $\approx\,65\,$nm. The line edge roughness (3\,$\sigma$) was evaluated to be less than 3\,nm both in the transfer layer and in silicon. We also demonstrate that a similar three-layer stack can be used for metal lift-off of high resolution patterns. A device application is demonstrated by fabricating 50\,nm half pitch dense nickel contacts to an InAs nanowire.
\end{abstract}


\maketitle 

\section{Introduction}
In Scanning Probe Lithography (SPL) the close proximity of the scanning tip to the substrate enables local surface manipulations down to atomic dimensions \cite{Garcia2014Advanced}. In particular for bias induced (b-SPL) and oxidation SPL (o-SPL) processes the achievable feature size is often smaller than the diameter of the scanning probe tip-apex because non-linear interactions result in a focussed interaction cross section. Prominent examples are the room temperature manipulation of single atoms in UHV environment \cite{sugimoto2005atom}, the creation of 3\,nm half-pitch parallel lines by field induced deposition of organic material \cite{martinez2007patterning}, or sub 10\,nm half-pitch patterning using local anodic oxidation (o-SPL) \cite{tello2002linewidth}. Thermal SPL (t-SPL) relies on the thermal decomposition of a thermally sensitive resist and has proven to be a high speed technique \cite{paul2011rapid} capable of writing 3D relief patterns in a single patterning step \cite{knoll2010probe}. In t-SPL 10\,nm half-pitch patterns were demonstrated having a patterning depth of 3-4\,nm \cite{cheong2013thermal}.

In particular for SPL but also for other lithographic techniques such as electron beam lithography (EBL) \cite{bonam2010performance}, electron beam induced deposition (EBID) \cite{van2011electron}, or He-Ion lithography \cite{sidorkin2009sub} highest resolution patterning is often accompanied by a low ($\lesssim\,10$nm) thickness of the produced features. Thus, further processing of the patterns for device fabrication is challenging and the number of publications is limited. With the aim of creating nanoimprint masters \cite{morecroft2009sub, peroz2012single} and bit-patterned magnetic media \cite{yang2011fabrication} EBL has been used to achieved sub 15\,nm half pitch gratings in 20-30\,nm thick HSQ resist. For o-SPL it was shown that a minimum oxide thickness of 1.1\,nm was required to fabricate 12\,nm thick and 30\,nm wide silicon nanowires \cite{ryu2014Fabrication}.

A method for amplifying shallow resist features has been developed in industry. It utilizes a three-layer hardmask stack which multiplies the pattern present in a thin photo-resist into a thick organic hardmask.\cite{bencher2011self} Recently, such stacks have been used to transfer self-assembled block copolymer patterns into a substrate using 10\,nm thick SiN hardmasks.\cite{bencher2011self, tsai2013pattern} In the case of t-SPL we addressed the pattern transfer issue by using a similar three-layer stack consisting of a polymer transfer layer covered by a sputter deposited 4\,nm silicon-oxide hard mask and the actual imaging resist. The latter may consist of a molecular glass material \cite{pires2010nanoscale} or the self-amplified-decomposition polymer polyphthalaldehye (PPA) \cite{cheong2013thermal}. For t-SPL the three layer stack serves two purposes. First, it thermally decouples the imaging resist from the underlying substrate which renders the writing process independent of the underlying substrate materials. Second, it provides amplification in the vertical direction by exploiting the different etch selectivities for SiO$_2$ and organic materials in reactive ion etch (RIE) etch processes. Recently we achieved a sub 30\,nm half-pitch pattern transfer into the silicon substrate \cite{cheong2013thermal} using a 6-8\,nm deep pattern in the 20\,nm thick imaging layer.

In this paper we describe our work to increase the resolution of t-SPL patterning and pattern transfer to below 20\,nm half pitch resolution. In particular we decreased the silicon oxide thickness to just 2-3\,nm and switched to an evaporation process for SiO$_2$ deposition. In addition we decreased the imaging transfer layer thickness to 9\,nm. We demonstrate that a $\approx 5\,$nm deep pattern with 18.3\,nm half pitch resolution can be transferred into the polymer transfer layer and subsequently into the silicon substrate. In addition we show that the transfer stack may also be used in a lift-off process if the highly cross-linked commercial HM8006 layer is replaced with PMMA. We demonstrate high resolution deposition of 25\,nm thick nickel into complex shaped structures. Finally we demonstrate a device application for the lift-off process by the fabrication of densely spaced metal contacts to an InAs nanowire.

\section{Pattern Transfer Stack}

\subsection{Materials, layer thicknesses and transfer steps}

 \begin{figure}[h]
 \includegraphics{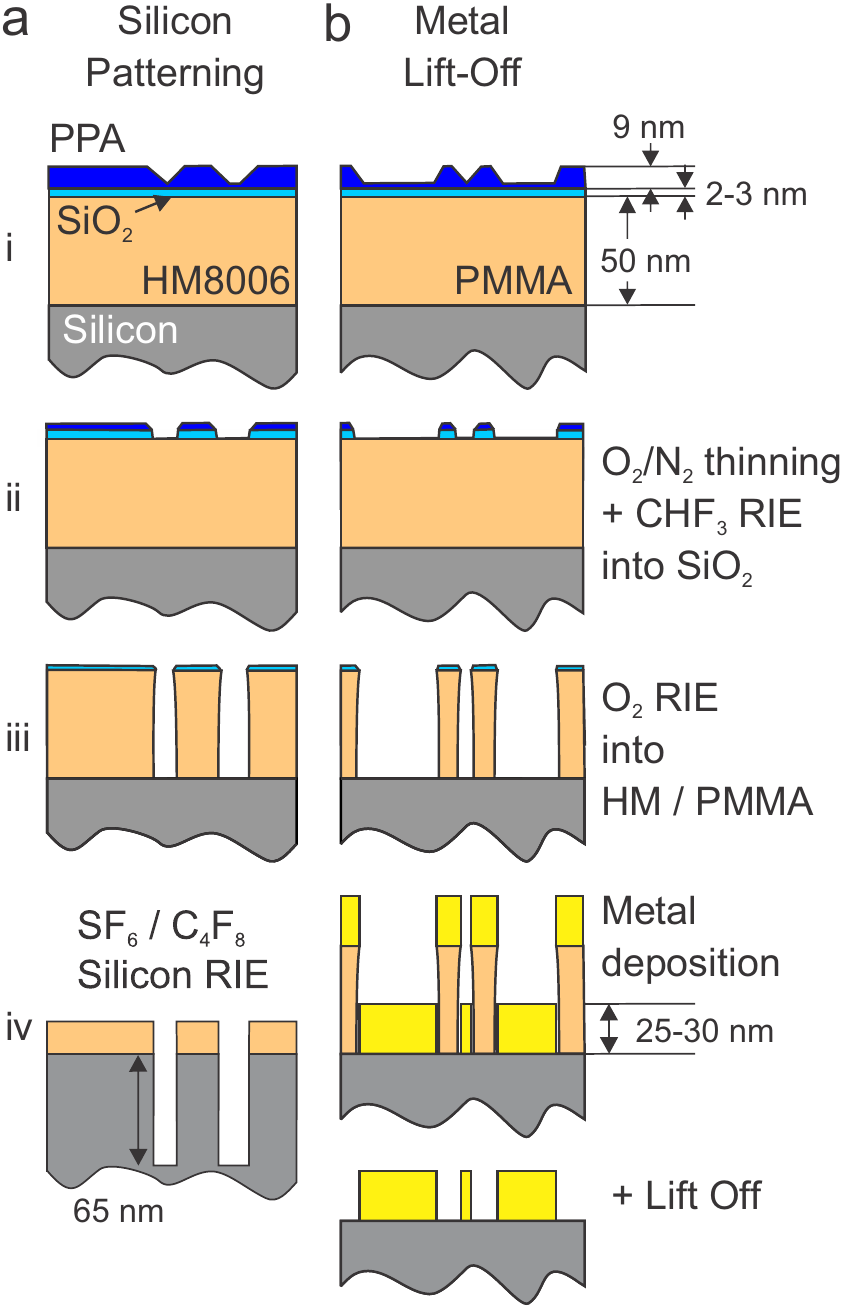}%
 \caption{\label{process} (Color online) Schematics of the pattern transfer process used for high resolution silicon pattern transfer (a) and metal lift-off (b). Step i: The patterning stack consists of a HM8006 or PMMA pattern transfer layer, a 2-3\,nm SiO$_2$ hard mask, and a $\approx\,9\,$nm PPA t-SPL resist layer. The patterning depth in t-SPL writing is 4 to 9\,nm into the PPA layer. Step ii: An O$_2$/N$_2$ PPA thinning and a CHF$_3$ RIE step is performed to transfer the pattern into the SiO$_2$ hard mask. Step iii: The pattern is etched into the HM8006 or PMMA layer by O$_2$ RIE for silicon transfer and metal lift-off, respectively. A slight anisotropic etch contribution creates some undercut of the lines. Step iv: In the silicon transfer step the pattern is processed to a depth of $\approx\,65\,$nm into silicon by deep RIE. For metal lift-off a 25\,nm thick metal layer is deposited. The PMMA matrix is removed in acetone.}%
 \end{figure}

The materials used in the transfer stack are optimized for low thermal conduction and high etch performance during pattern transfer. The transfer stack and the pattern transfer process flow are schematically depicted in Fig. \ref{process}. It exploits the high etch selectivity of the only several nm thick SiO$_2$ hard mask relative to the polymer layer in an oxygen RIE plasma. Thus any structure transferred into the SiO$_2$ layer may be readily amplified into a dual tone pattern in the transfer layer. One of the additional advantages of the three layer stack is the decoupling of the polymer layer functionality. The top layer is a pure imaging layer, which can be optimized for the t-SPL writing conditions. The lowest layer, called the transfer layer, may be optimized for the type of pattern transfer planned. For etch transfer into the substrate we use the highly cross-linked carbon rich HM8006 resist (JSR), which has a high mechanical stability and a good etch performance in fluorine containing plasmas. We chose a thickness of 50\,nm, compatible with current requirements for next generation lithography as defined in the ITRS roadmap for the year 2017 \cite{ITRS2013}. For the lift-off process we used a 53\,nm thick layer of PMMA, a material which is often used in EBL for lift-off processes.

As described in more detail below, the critical layer governing the performance of the transfer stack is the silicon oxide hard mask. Previously \cite{pires2010nanoscale,cheong2013thermal} a sputter deposited 4\,nm thick layer was used. Here we reduced that thickness to just 2-3\,nm and changed the deposition method to thermal evaporation (Pfeiffer PLS 500 Evaporation System). The change of the deposition method had a critical impact on the results shown here. First, the roughness of the SiO$_2$ is slightly improved from $R_a$ = 0.26\,nm to 0.23\,nm. Second, the sputter deposition of SiO$_2$ on PMMA resulted in a high roughness of the surface which is detrimental for the pattern transfer as well as for markerless overlay \cite{Rawlings2014}. Conversely thermal evaporation of SiO$_2$ on PMMA resulted in very smooth films, $R_a$ = 0.23\,nm. Third, on the sputtered oxide we had to use a minimal imaging resist thickness of 20\,nm in order to achieve a patterning depth of more than 5\,nm \cite{pires2010nanoscale,cheong2013thermal}. On the evaporated oxide the thickness of the PPA imaging resist could be reduced to 9\,nm. Furthermore it was possible to fully remove the PPA layer in t-SPL patterning, see for example the nanowire contact patterning at the end of this paper. The PPA layer was spin coated from cyclohexanone solution.

\begin{table*}
\caption{\label{table} Etch parameters for transfer process.}
\begin{tabular}{c c c c c c c c c}
\hline
Process & etched & rate & etch & rate & gases & power & pressure & etch time \\ [0.5ex]
 & layer & [nm/min] & mask & [nm/min] & & [W] & [mT] & [s]\\
\hline\hline
Silicon transfer& PPA & $10.5\pm1$ & & & O$_2$/N$_2$ 1:4 & 10 & 15 & 21 \\
 & SiO$_2$ & 14.0 & PPA & $5\pm2$ & CHF$_3$ & 100 & 15 & 14 \\
 & HM8006 & 19.5 & SiO$_2$ & negl. & O$_2$ & 20 & 15 & 169 \\
 & Si & 550 & HM8006 & $\approx 80$ & SF$_6$/C$_4$F$_8$ 1:1.5 & 1200 & 11 & 11 \\
\hline
Metal Lift-Off & PPA & $10.5\pm1$ & & & O$_2$/N$_2$ 1:4 & 10 & 15 & 34 \\
 & SiO$_2$ & 14.0 & PPA & $5\pm2$ & CHF$_3$ & 100 & 15 & 12 \\
 & PMMA & $34\pm2$ & SiO$_2$ & negl. & O$_2$ & 20 & 15 & 119 \\
\hline

\end{tabular}
\end{table*}

The stack, the process flow, and the etch conditions are similar to what has be described earlier \cite{cheong2013thermal}. A summary of the etch parameters is given in table \ref{table}. All etch processes were done using an Oxford Plasma Lab 80 tool except for the silicon etch which was done in an Alcatel AMS 200 deep RIE. For better performance of the critical transfer into the SiO$_2$ hard mask we did not remove the sample between the O$_2$/N$_2$ thinning step and the CHF$_3$ etch into the SiO$_2$ hard mask, which avoids exposure to ambient conditions. Thinning of the PPA was designed to reach a residual PPA thickness in the non-patterned areas of 3\,nm. Due to the $\gtrsim2$:1 SiO$_2$:PPA etch selectivity in the CHF$_3$ etch step this thickness is sufficient to ensure that the SiO$_2$ hard mask is not opened accidentally at an unwritten area.
In the last step the patterns defined in the SiO$_2$ hard mask are transferred either into HM8006 ready for etching into the substrate or into PMMA for lift-off of metal structures.

\subsection{Silicon Oxide Hard Mask Endurance}

\begin{figure}[h]
 \includegraphics{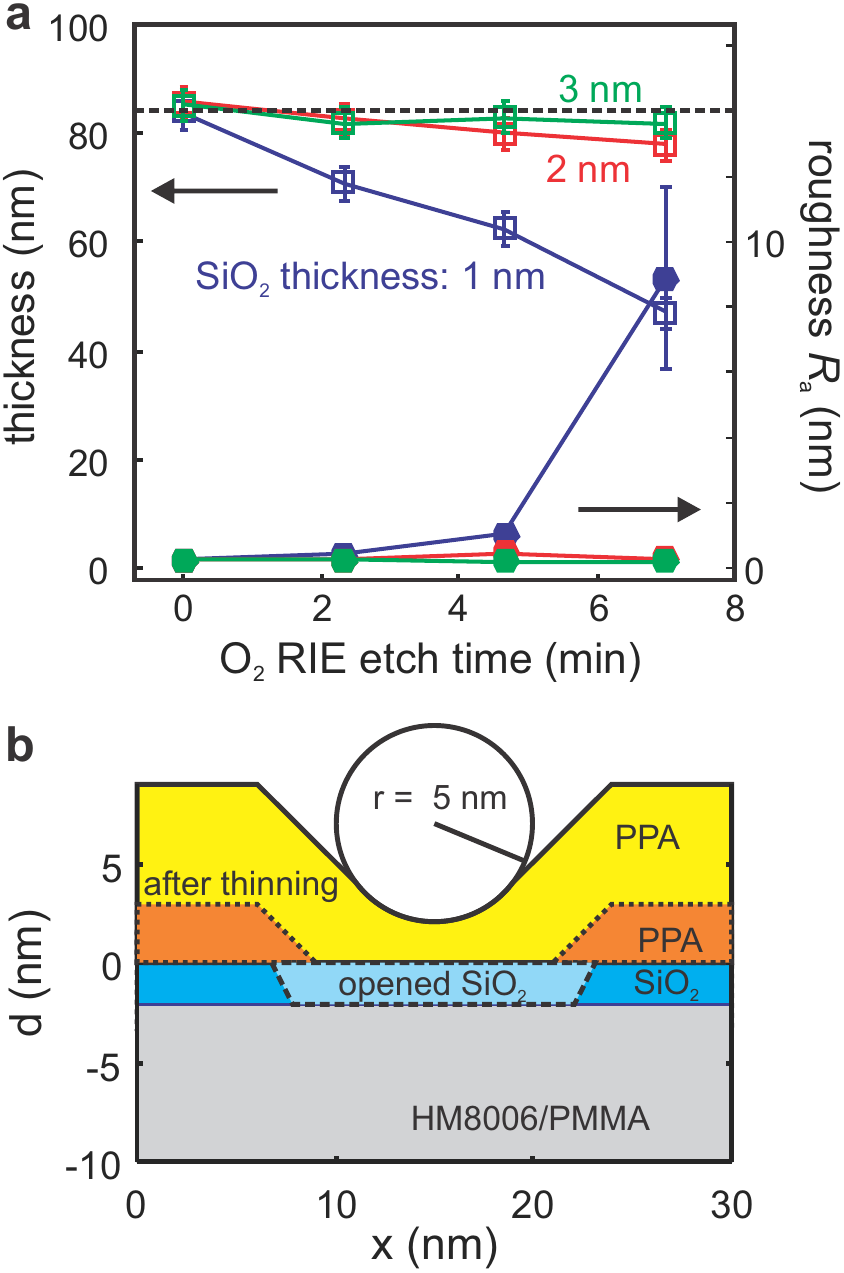}%
 \caption{\label{Etch} (Color online) Silicon hard mask optimization and schematics of the t-SPL patterning geometry. a) Remaining PMMA/SiO$_2$ layer thickness and surface roughness vs. O$_2$ RIE etch time for hard mask thicknesses of 1, 2, and 3\,nm. b) Schematic visualization of the t-SPL patterning geometry given the optimized SiO$_2$ layer thicknesses of 2\,nm. The initial shape in PPA (yellow) is given by the tip radius of 5\,nm. Thinning is designed to yield a 3\,nm thick remaining PPA layer (orange). The CHF$_3$ step for opening the SiO$_2$ hard mask (light blue indicates removed SiO$_2$) amplifies the slope of the profile by a factor corresponding to the etch selectivity of $\approx\,2$.}%
 \end{figure}

The silicon oxide hard mask is the critical layer in our pattern transfer strategy. It should be as thin as possible to allow for the transfer of shallow high resolution patterns in the PPA layer. At the same time it should be sufficiently thick to ensure a reliable transfer into the transfer layer at minimal SiO$_2$ thickness. In order to determine the optimal thickness of the SiO$_2$ hard mask we measured its stability during the oxygen pattern transfer step on PMMA, see Figure \ref{process}c. We deposited SiO$_2$ with thicknesses of 1, 2, and 3\,nm onto 83\,nm thick layers of PMMA and measured the residual thickness and the roughness of the surface as a function of etch time. The results are displayed in Figure \ref{Etch}a. Cumulative etch times of 140, 280 and 420s were used. The 1\,nm thick layer yields already after the first etch step as indicated by the reduced film thickness. Also the roughness rises significantly after the second etch step. For the 2 and 3\,nm thick layers the thickness and the roughness are constant within the experimental error, except for the last step in which the thickness of the 2\,nm stack is slightly reduced. Given the etch selectivities of the respective materials the 2\,nm SiO$_2$ membrane provides an etch depth of more than 90\,nm in HM8006 and more than 150\,nm of PMMA given an etch time of 280 seconds. Thus a 2\,nm thick membrane allows one to reliably transfer the SiO$_2$ pattern into the $\approx\,50\,$nm thick HM8006 or PMMA transfer layers.

The optimized value for the SiO$_2$ layer thicknesses of $2-3\,$nm is the basis for an optimal design of the imaging layer thickness. Given an etch selectivity of 1:2 (PPA: SiO$_2$), see table \ref{table}, the equivalent thickness of the transferred pattern in PPA is less than 1.5 nm. Adding the PPA surface roughness of $\approx 2\,$nm, a minimum patterning depth of 3-4 nm in PPA is required for a successful transfer. We found that a PPA thickness of $9\,$nm was sufficient to reliably yield patterns with 4-9 nm depth.

The final situation for t-SPL-writing and transfer into the SiO$_2$ hard mask is sketched in Figure \ref{Etch}b. If we assume a 5\,nm tip radius and a 45 degree opening angle of the written PPA patterns \cite{Holzner2011Directed,cheong2013thermal}, we arrive at a PPA profile as sketched in yellow. Thinning in the O$_2$/N$_2$ plasma leaves a residual PPA thickness of $\approx 3$\,nm (orange). Thinning does not affect the surface roughness significantly \cite{cheong2013thermal} and we assume a purely anisotropic etch. The most important aspect of the thinning procedure is to cut the t-SPL profile at a position where the slope is steepest, i.e. close to the PPA surface. As argued before\cite{cheong2013thermal}, a high slope is critical for obtaining small line edge roughness values. On the other hand three nanometer residual thickness is sufficient because of the low surface roughness of the PPA after the thinning of $R_a = 0.3\,$nm. We did not observe unintentional opening of the hard mask in non-written areas.

\subsection{Thermal Scanning Probe Lithography}

Writing is done as described elsewhere \cite{cheong2013thermal} in detail. We used a combined write and read approach where the writing is done in the trace direction of the scan and the reading is done in the retrace direction. At the end of the write/read process the topographical image was recorded. We used nominal tip heater temperatures of $\approx 800^\circ\,$C and force pulse durations of $5\,\mu$s. The linear scan speed varied between 0.2 and 0.3 mm/s depending on the write pixel size, which were either 6.9 or 9.2\,nm.

\section{Results and Discussion}

\subsection{18.3\,nm half pitch pattern transfer into silicon}

\begin{figure*}[ht]
\includegraphics{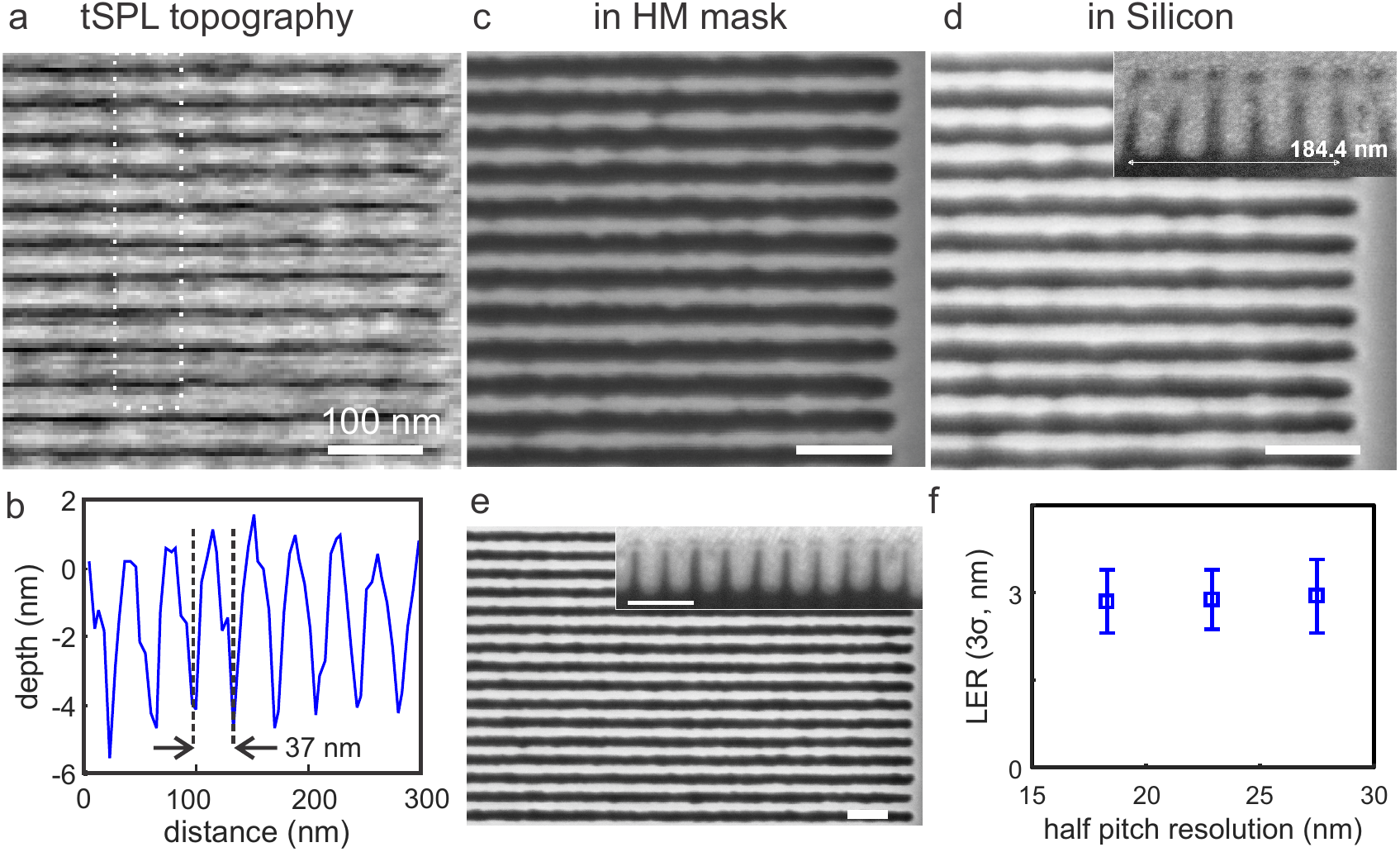}%
\caption{\label{Silicon} High resolution silicon patterning. a) Topography after t-SPL writing. The gray-scale corresponds to 7\,nm depth scale. b) Cross section according to the white dotted box in panel a. The patterning depth was 4 to 5\,nm at a full pitch of 36.6\,nm. c) SEM image after pattern transfer into the HM8006 transfer layer. d) SEM image after silicon transfer to a depth of $\approx\,65\,$nm. Note that the line roughness is correlated in images a, c, and d. e) SEM image of a 22.9\,nm half-pitch pattern in silicon. The insets in panels d and e show SEM cross-sections of the dense lines in similar fields after Pt deposition and FIB milling. f) Result of the line-edge roughness analysis of the SEM images after silicon transfer. The 3$\sigma$ value of the line edge roughness was $2.85\pm0.51$, $2.94\pm0.63$, and $3.24\pm0.66$\,nm$_{RMS}$ for pattern half pitches of 18.3, 22.9 and 27.5\,nm, respectively. All scale bars are 100\,nm. }%
\end{figure*}

Using the optimized transfer stack we studied the achievable resolution and line edge roughness (LER) for silicon patterning. We used the stack described above with a 50\,nm thick HM8006 layer for pattern transfer. The PPA thickness of 9.3\,nm was measured by AFM imaging of a scratch in a layer spin coated from the same solution onto a reference silicon wafer. Dense lines were written in the t-SPL tool with half pitch values of 18.3, 22.9 and 27.5\,nm using a pixel size of 9.2\,nm for writing. Imaging was done in the same tool using a pixel size of 4.5\,nm. Figure \ref{Silicon} depicts the results of the written structures in PPA, transferred into the HM8006 layer, and into the silicon substrate. The topography of the image recorded in the t-SPL tool is shown in Figure \ref{Silicon}a for the 18.3\,nm half pitch pattern. A cross-section perpendicular to the lines is shown in Figure \ref{Silicon}b. The topographical pattern is 4 to 5\,nm deep which was sufficient to transfer the pattern successfully through the SiO$_2$ hard mask and into the HM8006 transfer layer. Figure \ref{Silicon}c shows a scanning electron microscopy (SEM) micrograph of the pattern etched into the HM8006 layer. A final SF$_6$/C$_4$F$_8$ etch step was used to transfer the pattern to a depth of $\approx\,65\,$nm into the silicon wafer, see SEM micrograph in Figure \ref{Silicon}d. The final pattern in silicon with a half pitch value of 22.9\,nm is shown in Figure \ref{Silicon}e. SEM cross-sectional images of the dense lines in similar fields were recorded after Pt deposition and Focused Ion Beam (FIB) milling using a FEI Helios 450S, see insets in Figure \ref{Silicon}d and e.

For all half-pitch values the line edge roughness (LER) in the SEM images was determined by analyzing contour-lines for a given image intensity value. The standard deviation of the contour lines was calculated after subtracting a common linear fit for all contour lines to account for the rotation of the image. The values for the LER are defined by three times the standard deviation ($3\,\sigma$) and are shown in Figure \ref{Silicon}f. The error-bars in the plot denote the standard deviation of the LER as obtained for the individual contour lines.

The obtained value for the $3\,\sigma$ line edge roughness of $2.9\,\pm\,0.5$\,nm is in agreement with the results obtained previously for 27.5\,nm lines and spaces\cite{cheong2013thermal}. In this previous work we found that the LER was inversely proportional to the slope of the topographic patterns. Here the pattern density is increased by a factor of 1/3 and the depth is reduced by a factor of 1/3, which suggests that the slope is similar in both cases. Therefore we expect similar LER values, which is indeed the case. The results obtained here therefore corroborate the assumption that the line edge roughness in t-SPL originates from the roughnesses of the participating interfaces, e.g. the PPA and the SiO$_2$ roughness, and the slope of the written patterns. As a consequence scaling to higher line densities implies new approaches are required to optimize these values.

\subsection{High resolution metal lift-off}

\begin{figure}[h]
 \includegraphics{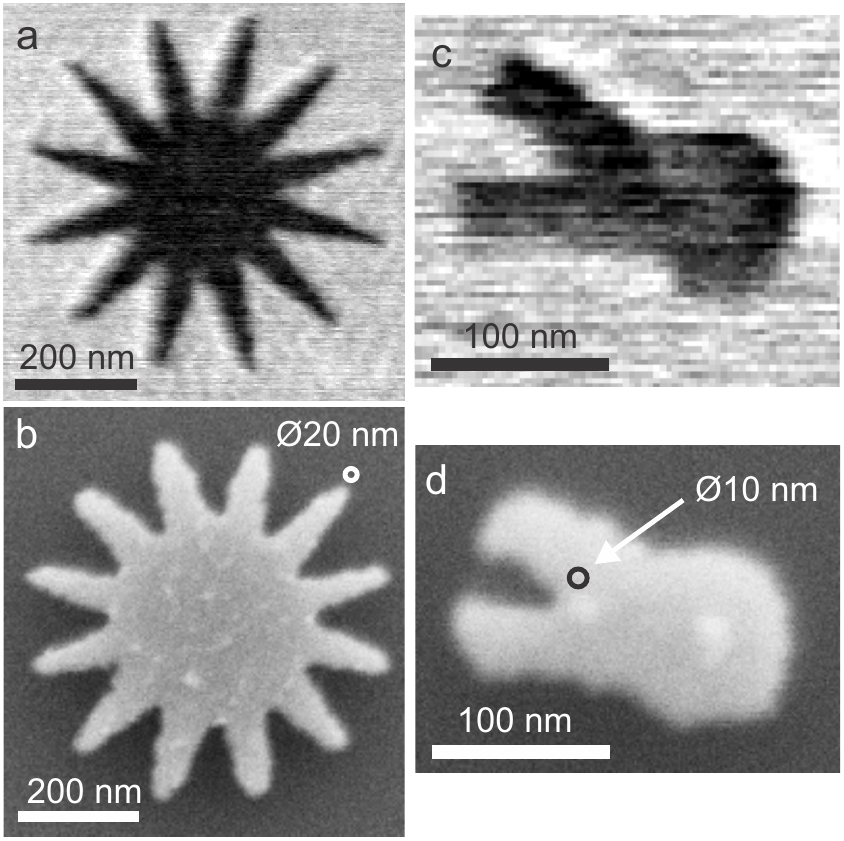}%
 \caption{\label{Lift-off} Demonstration of the metal lift-off process. a) and c) Topography images recorded in the t-SPL tool. b) and d) Respective patterns after 25\,nm nickel lift-off on silicon. Minimal feature sizes can be estimated by comparison with the displayed circles of diameter 10 and 20\,nm.}%
 \end{figure}

Using the three layer transfer stack for patterning effectively decouples the functions of the individual layers. Thus the bottom transfer layer can be optimized for a different purpose, here lift-off deposition of metal layers. HM8006 is not suited for lift-off purposes due to the highly cross-linked nature of this material. The covalent cross-links cannot be opened by solvents which impairs removal of the layer. For lift-off a mechanically stable linear polymer should be used which can be easily removed by solvents. We chose to use PMMA because it has a high mechanical stability and is typically used in the EBL environment.

A second important requirement for a successful lift-off process is a geometrical undercut in the patterned structures. After opening the transfer layer using the silicon oxide hardmask such an undercut indeed exists as schematically depicted in Figure \ref{process}b (iii). The undercut originates from a slightly isotropic component of the oxygen RIE etch which widens the trenches below the hard mask by a few nanometers. Thus, as depicted in Figure \ref{process}d, the shape of the pattern in the transfer layer provides the necessary decoupling of the metal layer deposited on top of the remaining structures to the metal layer at the bottom of the substrate.

In order to demonstrate the quality of the obtained lift-off process we chose a PMMA layer thickness of 53\,nm. After etching into this layer a 25\,nm thick nickel film was evaporated using electron-beam evaporation (Evatec). The working distance was 600\,mm and the evaporation pressure below $10^{-6}$\,mbar. The lift-off was done in acetone followed by cleaning the wafer in isopropanol without applying ultrasound. The result of the process is shown in Figure \ref{Lift-off} for two exemplary high resolution patterns. A first example is a star-shaped pattern of 500\,nm diameter with 150\,nm long and less than 30\,nm thin arms. The topographic pattern shown in Figure \ref{Lift-off}a has a medium depth of 5\,nm. The lift-off pattern shows high resolution features as indicated by the 20\,nm circle drawn next to the structure. The direct comparison of the two images reveals that the metal structure is slightly wider than the t-SPL written structure which may be due to some diffusion or broadening during the metal deposition. The second structure shown in \ref{Lift-off}c and d is a circular shape with two handles. The length of the entire structure is about 200\,nm, the width of the handles less than 50\,nm. The opening between the handles is well reproduced in the lift-off pattern and exhibits an apex diameter of about 10\,nm as indicated in the figure by the added circle.

Writing such complex high resolution patterns in a t-SPL tool is straightforward since the writing process is not impaired by proximity effects. In t-SPL the pattern is created pixel by pixel from a chemical reaction decomposing the imaging resist PPA into volatile monomers. After each pixel the tip is removed from the surface and the temperature in the resist relaxes back to room temperature much faster than the arrival of the next patterning event. Therefore proximity effects are absent in this patterning method and complex shapes can be obtained using the same patterning conditions as for dense lines. Furthermore due to the thermal decoupling from the substrate by the transfer stack, patterning performance does not depend on the materials present at the substrate. We demonstrate this fact by fabricating closely spaced contacts to an InAs nanowire.

\subsection{InAs nanowire contacts}

The lift-off process described in this paper was applied to the technologically relevant problem of overlaying metal electrodes onto an InAs nanowire. The t-SPL patterning step produced an electrical connection between an existing lower resolution e-beam pattern and the nanowire. This enabled a fundamental investigation of the nanowire's transport properties which will be detailed elsewhere. t-SPL has sub-nanometer sensitivity to topography and can read a surface without causing resist exposure. This enables a markerless approach to overlay. We have shown that nanometer accurate pattern overlay can be achieved using this approach \cite{Rawlings2014}.

\begin{figure}

\includegraphics{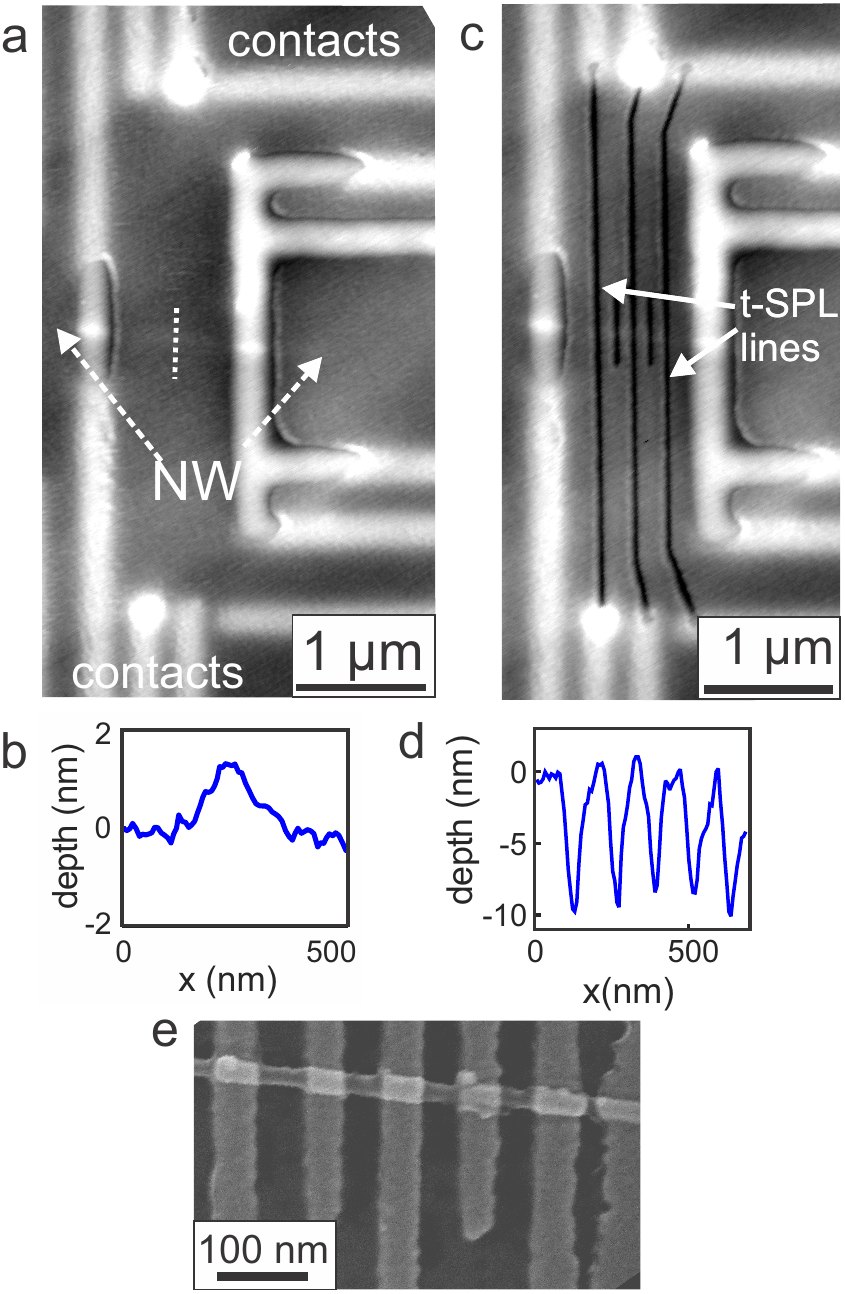}%
\caption{\label{Nanowire} Precise deposition of metal electrodes. a) Topography recorded before patterning. An InAs nanowire (NW) was deposited on silicon and contacted using EBL and lift-off. After deposition of the transfer stack the topography was recorded. b) Cross section according to the white dotted line in a). A residual topography of 1.5\,nm is measured. c) Result of t-SPL writing at the NW position. d) Cross-section of the written pattern along the NW axis. The patterning depth is $\approx 9$\,nm, corresponding to the PPA thickness. d) SEM micrograph of the finalized metal electrodes.}%
\end{figure}

The topography prior to t-SPL patterning measured is shown in Figure \ref{Nanowire}a. 30\,nm diameter nanowires were dispersed onto a silicon substrate which contained reference marks from a preceeding optical lithography step. The wire position was then determined with respect to these reference marks using an optical microscope and gold electrodes with a 250\,nm $\times$ 40\,nm cross-section were placed around the nanowire. Next the t-SPL lift-off patterning stack was added to the sample to yield the surface topography shown in Figure \ref{Nanowire}a. This topography was measured using the t-SPL tool and although the image contrast is dominated by the 15\,nm residual topography of the buried e-beam features the nanowire is still readily visible (Figure \ref{Nanowire}b).

The vector graphics package \textsc{Inkscape} was used to define the desired t-SPL electrode pattern with respect to the measured topography. The overlay code\cite{Rawlings2014} was then used to write this pattern in registry with the measured topography to yield the topography shown in Figure \ref{Nanowire}c. The write pixel size was 14\,nm and the nominal width of the wire was set to 28\,nm. The PPA layer was selectively removed to a depth of nearly 9\,nm (see Figure \ref{Nanowire}d), which corresponds to the nominal thickness of the PPA layer. Figure \ref{Nanowire}e shows an SEM image of the device in the vicinity of the nanowire following the lift-off process. Both the 30\,nm diameter nanowire and the 50\,nm wide metal electrodes defined in the t-SPL step are visible.

\section{Conclusion}

In conclusion we have shown that t-SPL is a viable technology for the creation of sub 20\,nm structures in a substrate and in metal deposited patterns. In comparison to previous results the high resolution was achieved by optimization of the pattern transfer stack. Since high resolution in t-SPL is linked to shallow patterns the stack was optimized to reliably process shallow resist patterns of less than 5\,nm depth. The shallow patterns are amplified by exploiting highly selective etch processes involving a very thin silicon oxide membrane. We determined the minimal thickness of this mask to be $\approx2\,$nm for a successful transfer into a 50\,nm thick transfer polymer. Consequently the thickness of the imaging resist PPA could be reduced to 9\,nm and high resolution patterning of 18.3\,nm dense lines with a LER value of less than 3\,nm was achieved. Deposition of the SiO$_2$ by thermal evaporation also allowed us to exchange the transfer layer polymer HM8006 by PMMA and perform a high resolution metal lift off on silicon and for contacting an InAs nanowire. Both examples demonstrate the ability of t-SPL to fabricate complex high resolution shapes independent of the nature of the substrate. Proximity effects are absent and the same writing conditions can be used for all examples demonstrated. Writing contacts to the nanowire demonstrates another strength of t-SPL which is the sub-nanometer topographical sensing capability allowing for a nanometer precise markerless overlay process.

\section{Acknowledgements}

The authors thank Ute Drechsler for the fabrication of the cantilevers and assistance in the RIE etch steps, Meinrad Tschudy for the deposition of the silicon oxide hard masks, Steffen Reidt for obtaining the cross-sectional images, Sigi Karg, Felix Holzner and Philip Paul for valuable discussions. The work was partially supported by the European Commission FP7-ICT-2011 no. 318804 and by the European Research Council StG no. 307079.


%
%


\providecommand{\noopsort}[1]{}\providecommand{\singleletter}[1]{#1}%

\end{document}